\documentclass[aps,prl,reprint,groupedaddress,floatfix]{revtex4-2}

\usepackage[printonlyused]{acronym}
\usepackage{graphicx}
\usepackage{subcaption}
\usepackage{dcolumn}
\usepackage{bm}
\usepackage{makecell}
\usepackage{rotating}
\usepackage{orcidlink}
\usepackage{tabularx}
\usepackage{booktabs}
\usepackage{braket}
\usepackage{framed}
\hypersetup{breaklinks=true,hidelinks}
\usepackage[english]{babel}

\begin{document}

\title{Learning quantum properties with informationally redundant external representations: An eye-tracking study}

\author{Eva Rexigel}
\thanks{These authors contributed equally to this work.}
\affiliation{Department of Physics and State Research Center OPTIMAS, University of Kaiserslautern-Landau, Erwin-Schroedinger-Str. 46, 67663 Kaiserslautern, Germany}

\author{Linda Qerimi}
\thanks{These authors contributed equally to this work.}
\affiliation{Chair of Physics Education, Faculty of Physics, Ludwig-Maximilians-Universität München, Geschwister-Scholl-Platz 1, 80539 Munich, Germany, and \\Munich Quantum Valley (MQV), Max Planck Institute of Quantum Optics (MPQ), Germany}

\author{Jonas Bley}
\affiliation{Department of Physics and State Research Center OPTIMAS, University of Kaiserslautern - Landau, Erwin-Schroedinger-Str. 46, 67663 Kaiserslautern, Germany}

\author{Sarah Malone}
\affiliation{Department of Education, Saarland University,  Campus A4 2, 66123 Saarbrücken, Germany}

\author{Stefan Küchemann}
\affiliation{Chair of Physics Education, Faculty of Physics, Ludwig-Maximilians-Universität München, Geschwister-Scholl-Platz 1, 80539 Munich, Germany}

 \author{Jochen Kuhn}
\affiliation{Chair of Physics Education, Faculty of Physics, Ludwig-Maximilians-Universität München, Geschwister-Scholl-Platz 1, 80539 Munich, Germany}

\begin{abstract}
Recent research indicates that the use of \ac{mer} has the potential to support learning, especially in complex scientific areas, such as quantum physics. In particular, the provision of informationally redundant external representations can have advantageous effects on learning outcomes. This is of special relevance for quantum education, where various external representations are available and their effective use is recognised as crucial to student learning. However, research on the effects of informationally redundant external representations in quantum learning is limited. The present study aims to contribute to the development of effective learning materials by investigating the effects of informationally redundant external representations on students' learning of quantum physics. Using a between-subjects design, 113 students were randomly assigned to one of four conditions. The control group learnt with a traditional multimedia learning unit on the behaviour of a single photon in a Mach-Zehnder interferometer. The three intervention groups received redundant essential information in the Dirac formalism, the Bloch sphere, or both. The use of eye tracking enabled insight into the learning process depending on the external representations provided. While the results indicate no effect of the study condition on learning outcomes (content knowledge and cognitive load), the analysis of visual behaviour reveals decreased learning efficiency with the addition of the Bloch sphere to the multimedia learning unit. The results are discussed based on current insight in learning with \ac{mer}. The study emphasises the need for careful instructional design to balance the associated cognitive load when learning with informationally redundant external representations.

\end{abstract}


\maketitle

\section{Introduction}\label{sec:intro}
\subsection{Background and Motivation}

Science education, particularly in the domain of physics, is characterised by the effective use of external representations.
These may include textual descriptions, equations and formulas, diagrams, and graphs, or educators' explanations.
This is especially the case for complex physics concepts, such as those encountered in the context of quantum physics, a field characterised by its abstract principles and counterintuitive phenomena.
In such cases, external representations play a crucial role in the communication and education of the subject \cite{singh_review_2015}.
It has been shown that quantum education based on classical analogies often leads to conceptual difficulties \cite{Bouchee.2022}.
Consequently, the judicious use of external representations in quantum physics education is essential to prevent misconceptions and facilitate a more profound understanding of quantum phenomena.
Across a range of science, technology, engineering and mathematics (\ac{stem}) disciplines, the use of multiple external representations (\ac{mer}) has been evidenced as an effective tool for fostering student learning (for an overview, see \cite{ainsworth_deft_2006, rexigel_investigating_2024}).
This is particularly the case in contexts characterised by high complexity \cite{Ainsworth.2008}.
Consequently, it may also prove to be a valuable method for assisting students in the effective acquisition of quantum concepts.

Understanding key concepts in quantum physics has become increasingly important in recent decades as the relevance of quantum technologies has grown \cite{merzel_core_2024}.
With its rapidly developing pillars of quantum communication, quantum computation, quantum simulation, and quantum sensing, a particular focus is placed on two-state systems, commonly referred to as qubits.
The introduction of a qubit and its quantum properties has been demonstrated to be an effective method to introduce the fundamental principles of quantum physics and quantum technologies \cite{sadaghiani_spin_2015, dur_was_2012}.
The behaviour of a single photon in a Mach-Zehnder interferometer (\ac{mzi}) is a common experimental approach that provides a valuable foundation to understand fundamental quantum concepts \cite{marshman_investigating_2017}.
Studies show that the \ac{mzi} is a helpful tool to reduce comprehension difficulties and improve students' understanding of wave-particle duality and the probabilistic nature of quantum measurement, as it demonstrates the principles of quantum mechanics in a tangible experimental setting \cite{marshman_investigating_2017}.
Despite the introduction of a variety of teaching strategies in quantum physics in recent years \cite{PhysRevPhysEducRes.20.020601}, quantum physics concepts continue to present a considerable challenge to learners across different levels of education and academic backgrounds \cite{Bitzenbauer.2022, hennig_introducing_2024}.
Indeed, previous research indicates that developing a comprehensive understanding of quantum physics requires a substantial shift in perspective, diverging from classical concepts, which often leads to misconceptions \cite{singh_review_2015}.

The use of \ac{mer} with shared information enables different representations of the same essential information. It is important to emphasize that informational redundancy in this context refers specifically to the essential information aligned with the learning objective. For example, the understanding that the general state of a qubit is a superposition of two basis states can be conveyed either through Dirac notation or via a graphical representation such as the Bloch sphere. Both external representations hold the essential information that the general state of a qubit is a superposition of two basis states, yet they differ in the additional information they convey and the manner in which it is presented. These differences provide learners with complementary perspectives on the concept and can support learning by engaging distinct cognitive processes \cite{Ainsworth.2021}.

The aim of this study is to investigate the potential of \ac{mer}, particularly those that are informationally redundant, to facilitate the learning of fundamental quantum properties, illustrated by the single-photon behaviour in a \ac{mzi}.

\subsection{Learning with MERs}\label{subsec:mer}

The acquisition of scientific knowledge is contingent on the use of suitable external representations. These serve as the foundations for effective communication, allowing us to convey information in a multitude of formats tailored to the specific requirements of the situation. It is generally accepted that there is a distinction to be made between symbolic representations, encompassing text, equation, and formula, and graphical representations, which include, for example,  diagram and graph \cite{schnotz_kognitive_2001, Schnotz.2021}. In contrast to symbolic external representations, which are based on symbols that bear no direct resemblance to the referent, graphical external representations are based on icons that share structural characteristics with the referent, such as similarity in shape or form \cite{Schnotz.2021}. 

Current research indicates that the use of \ac{mer} has the potential to facilitate learning in different \ac{stem} contexts, in contrast to the use of a single external representation cite{Ainsworth.2021}.
In this context, a notable focus has been on the advantages of learning with text and pictures, known as multimedia learning, compared to learning through text alone \cite{Mayer.2021g}.
The beneficial effect of combining text and pictures, as opposed to text alone, is commonly referred to as the multimedia principle \cite{Mayer.2021c}.
According to cognitive theories such as the \ac{ctml} and the \ac{itpc}, the multimedia principle can be explained by a more efficient use of cognitive resources due to the dual structure of sensory memory and working memory, which allows for parallel processing of symbolic and graphical structures \cite{Mayer.2021e, Schnotz.2021}.
In line with the \ac{ctml}, the cognitive process of learning with \ac{mer} consists of three fundamental stages, including selection, organisation, and integration processes \cite{Mayer.2021e}.
First, learners must select relevant information encoded in the external representations provided.
Second, they need to organise the relevant information into mental structures. 
Third, learners must use these mental structures to build a comprehensive mental model by combining them with existing knowledge retrieved from long-term memory. 
The benefits of \ac{mer} have also been identified for various combinations of symbolic and graphical representations \cite{rexigel_more_2024}.
In particular, recent research has shown that the advantages of \ac{mer} are not limited to heterogeneous combinations of symbolic and graphical external representations.
In fact, they can also be detected at a similar level in homogeneous combinations of multiple symbolic external representations \cite{ott_multiple_2018, malone_homogeneous_2020}.
The \ac{itpc}, developed by Schnotz and Bannert, complements this perspective by highlighting the importance of semantic coherence between representations \cite{schnotz2003construction}.
According to the \ac{itpc}, graphical representations only have the potential to facilitate learning only if they are semantically aligned with the accompanying symbolic representation(s) and do not contain any contradictory information \cite{schnotz2005integrated, schnotz2001kognitive}.

In addition to the cognitive theories of multimedia learning, the \ac{deft} framework defines three main functions that \ac{mer} can fulfil to support learning \cite{ainsworth_deft_2006,Ainsworth.2021}.
Regardless of the specific types of external representations combined, \ac{mer} can facilitate learning by complementing each other, constraining each other, or constructing a deeper understanding \cite{Ainsworth.2021}.
In doing so, external representations can complement each other, either through information or through cognitive processes induced by the different representation of information.
They can constrain cognitive processing by focusing attention on relevant aspects.
Finally, they can construct deeper understanding by allowing learners to integrate information from different sources of information \cite{Ainsworth.2021}.

The \ac{deft} framework provides explanations for the learning effectiveness of various combinations of external representations, particularly for learning with informationally redundant representations.
Providing \ac{mer} with shared information has the potential to support learners by inducing different cognitive processes and thus providing different access to the essential information \cite{Ainsworth.2021}.
In their recent meta-analysis, \cite{rexigel_more_2024} found that the provision of additional informationally redundant external representations has the potential to help students use cognitive resources more efficiently without providing additional essential information.
As a possible explanation for the beneficial effects of a higher number of \ac{mer} with shared information, the authors suggest that additional informationally redundant external representations increase the options for choosing the most appropriate external representation \cite{rexigel_more_2024}.
However, in order to benefit from multiple sources of the same information, learners need representational competence \cite{rau_conditions_2017, Daniel.2018b}.
According to \cite{rau_conditions_2017} representational competence covers three areas of expertise.
First, conceptual competencies are needed, including visual understanding of each external representation and connectional understanding of how the representations relate to each other.
Second, learners need perceptual competencies to be able to apply visual and connectional understanding fluently.
The third area of competence is given by meta-representational competencies, including the ability to choose an appropriate external representation based on the learning setting and personal characteristics \cite{rau_conditions_2017}. 

Despite the potential advantages of \ac{mer} with shared information, previous research has also revealed instances where the provision of multiple informationally redundant representations hinders learning.
According to the redundancy principle in its traditional form, learning with pictures and spoken text is more beneficial to learning than the additional presentation of printed text \cite{Fiorella.2021b}. 
Based on the most prominent version of the \ac{clt}, cognitive load when learning can be categorised in extraneous cognitive load (\ac{ecl}), intrinsic cognitive load (\ac{icl}), and germane cognitive load (\ac{gcl}).
Extraneous cognitive load is the result of the learner's interaction with elements introduced by the instructional design and should be reduced when learning with \ac{mer} to support learning \cite{Paas.2021}.
In contrast, \ac{icl} is the result of the learner's interaction with those elements that are intrinsic to the task and must be processed in parallel.
Finally, \ac{gcl} is determined by the amount of cognitive resources allocated to \ac{icl} rather than \ac{ecl} \cite{Paas.2021}.
Various approaches exist for measuring cognitive load. However, in the context of multimedia learning, subjective rating scales are most commonly used \cite{MutluBayraktar.2019}. Although such scales are influenced by retrospective self-assessment and individual self-concept \cite{Sweller2011}, several instruments have been developed and validated in recent years to provide reliable instruments for assessing \ac{ecl}, \ac{icl} and \ac{gcl} separately in diverse educational settings (for an overview, see \cite{MutluBayraktar.2019}).

The \ac{clt} provides an explanatory approach for the redundancy principle.
Each external representation provided to learners constitutes an additional source of information that needs to be processed and coordinated, resulting in an increase in \ac{ecl} \cite{Kalyuga.2021}.
In line with this, avoiding informationally redundant external representations frees cognitive resources for learning \cite{Kalyuga.2021}.
However, it has been shown that the learner characteristics play an important role in the effectiveness of redundant external representations \cite{Kalyuga.2021.expertise-reversal}. 
While the presentation of redundant information in different forms may be valuable for novices in providing different accesses to the relevant information, this advantage may diminish with increasing expertise. This is because the additional external presentation does not add value, but only increases \ac{ecl}.
This constitutes the expertise-reversal principle \cite{Kalyuga2010, Kalyuga.2021.expertise-reversal}.

Thus, previous research both supports advantageous effects of learning with multiple informationally redundant external representations \cite{Ainsworth.2021,rexigel_more_2024} and disadvantageous effects \cite{Kalyuga.2021}.

\subsection{The relevance of Dirac notation and the Bloch sphere in Quantum Education}

Especially in the field of quantum technology education the Bloch sphere and Dirac notation have been identified as external representations with high relevance \cite{rexigel_investigating_2024}. In educational contexts, conceptual advantages of the Dirac notation were recently discussed \cite{PhysRevPhysEducRes.20.020134}, with the results suggesting that the use of the Dirac notation facilitates the sensemaking of mathematics (probability rule, superpositions, orthogonality) and physics (connection to phenomena such as polarisation, measurements, and wave functions) and therefore acts as a bridge between mathematical structures and physical phenomena. The use of the Dirac notation has been shown to facilitate the understanding of intricate concepts in quantum mechanics \cite{wawro_students_2020, merzel_mathematical_2024}. In particular, the Dirac notation provides a concise representation of eigenvalues and eigenstates, establishing a strong connection between mathematical and physical concepts. 

While symbolic representations are often used in quantum education to formally explain quantum phenomena, graphical representations, such as the Bloch sphere, provide a vivid way to visualise and facilitate the understanding of quantum states \cite[e.g.,][]{hu_student_2024}.
However, previous research has also revealed some difficulties in learning with the Bloch sphere.
For example, students were found to have learning difficulties in constructing Bloch sphere states, understanding relative and global phases, and describing measurements when learning with the Bloch sphere \cite{hu_student_2024}.
Every dynamic of a quantum state can be interpreted in the Bloch sphere as a rotation of the state vector.
When dynamic content is presented statically, learners have to perform cognitively demanding mental transformations that are closely related to spatial visualisation skills \cite{hoffler2011role}.
While learners with high spatial competencies are capable of executing such processes mentally, learners with low spatial competencies benefit more from external animations \cite{hoffler2011role}.
Consequently, learners with higher spatial competences, in particular those with superior mental rotation skills, may benefit more from the Bloch sphere than those with less developed mental rotation ability.
Tests such as the RCube-Vis test \cite{Fehringer.2020} provide a differentiated measure of individual differences in mental rotation ability, while minimising the influence of other visual processing factors.
In the RCube-Vis test, two static representations of a Rubik's Cube are presented simultaneously, one in a rotated position and the other solved. The participant has to decide whether the presented cubes can be transformed into each other.
Similar to the rotation of the Bloch vector within the Bloch sphere, the individual layers of the cube must be mentally rotated.

\subsection{Interaction of visual and cognitive processes in learning with MERs}

The use of eye-tracking technology has proven to be a valuable tool in gaining insight into cognitive processing when learning with \ac{mer} \cite{Alemdag.2018,Coskun.2022}. For example, Klein et al. (2020) found that eye tracking provides valuable insight into the cognitive processes involved in graph comprehension, revealing different visual attention patterns when students solve kinematics problems depending on their response accuracy and confidence \cite{klein2020visual}. According to the systematic review by Hahn and Klein (2022), the analysis of gaze transitions also provides valuable insight into how learners integrate different sources of information, revealing differences in cognitive processing and problem-solving strategies \cite{hahn2022eye}. For instance, the number of transitions, defined as gaze shifts between defined areas of interest, such as different external representations, is a commonly used measure of learners' integration processes \cite{Alemdag.2018}.
Current research suggests that the frequency of transitions reflects the degree of cognitive interplay between text and visualisations, with more transitions indicating active efforts to connect both sources \cite[e.g.,][]{schmidt2010closer}.
Canham and Hegarty (2010) showed that learners with higher prior knowledge focus their transitions on task-relevant features \cite{rasch2009interactive}, while those with less knowledge may allocate their attention inefficiently.
Similarly, Hannus and Hyönä (1999) found that high-achieving students made more targeted transitions between text and illustrations in science textbooks than low-achieving students \cite{hannus1999utilization}, highlighting the importance of deliberate gaze shifts for effective comprehension.
In addition, transitions can be influenced by design features. Visually salient or cued elements tend to attract attention and promote smoother transitions between different components of the material \cite{hyona_use_2010}. 

\subsection{Research Questions}\label{subsec:rq}
Learning quantum physics is particularly challenging due to its abstract and counterintuitive nature.
Current research suggests that the use of \ac{mer} with shared information may be an effective way of supporting learning through a more efficient use of cognitive resources compared to learning with a single one \cite{rexigel_more_2024}.
However, it is not clear whether integration processes are responsible for this advantage or the fact that learners have the opportunity to choose the most appropriate external representation as opposed to learning with an individual representation.
For example, \cite{ott_multiple_2018} showed that the number of transitions between heterogeneous combinations of text and picture was higher than the number of transitions between homogeneous symbolic combinations of text and equation.
This could suggest that in the case of heterogeneous combinations of symbolic and graphical representations, integration processes are more likely to provide advantages of \ac{mer} and, in the case of homogeneous combinations, the possibility of choosing an appropriate one.
In light of the previous considerations, we investigate three research questions:

\begin{description}
    
    \item[\textup{\textbf{RQ1}}] Does adding an information-redundant symbolic-mathematical or graphical geometric representation to a multimedia learning unit enhance learning (content knowledge and cognitive load) of quantum properties?
    
    \item[\textup{\textbf{RQ2}}] Does the integration of both informationally redundant representations additionally promote learning?

    \item[\textup{\textbf{RQ3}}] Are advantages in learning with information-redundant representations correlated with visual integration processes across representations or rather the selection of one preferred representation?

\end{description}

This study was preregistered on the Open Science Framework (OSF) to ensure transparency and rigour \cite{Rexigel.2024c}.

\section{Methods}\label{sec:methods}
 
\subsection{Participants}\label{subsec:participants}
A total of 113 students from three German universities (RPTU Kaiserslautern-Landau, Ludwig-Maximilians-Universität München, and Saarland University) participated in the study. Participants were selected from a variety of fields related to \ac{stem} and randomly assigned to one of four groups: the control group ($N=28$), the intervention group IG1 ($N=28$), the intervention group IG2 ($N=28$) or the intervention group IG3 ($N=29$). A detailed overview of the number of participants in each group according to the field of study can be found in Table \ref{tab:descriptives.groups}. Three participants did not specify their field of study. In total, 71 men and 40 women were involved in the study. Two participants declined to specify their gender.

\begin{table}[htbp]
    \renewcommand{\arraystretch}{1.2}
    \begin{tabularx}{\linewidth}{>{\arraybackslash}X>{\centering\arraybackslash}X>{\centering\arraybackslash}X>
    {\centering\arraybackslash}X>{\centering\arraybackslash}X}
        \hline
        \textbf{Discipline} &\textbf{CG}& \textbf{IG1} & \textbf{IG2} & \textbf{IG3} \\
        \hline
        Physics&18 & 17 & 15 & 16 \\
        Mathematics& 1 & 1 & 0 & 2 \\
        
        Biology& 2 & 1 & 4 & 3 \\
        %\hline
        Biophysics  & 1 & 0 & 0 & 1 \\
        %\hline
        Business  & 1 & 0 & 0 & 0 \\
        %\hline
        Chemistry& 0 & 1 & 1 & 0 \\
        %\hline
        Education& 0 & 6 & 4 & 7 \\
        %\hline
        Engineering& 3 & 1 & 1 & 0 \\
        %\hline
 
        %\hline
        Pharmacy& 1 & 1 & 0 & 0 \\
        %\hline
        %\hline
        NA&1 & 0 & 2 & 0 \\
        
        \hline
    \end{tabularx}
    \caption{Number of participants in each of the four study conditions (CG, IG1, IG2, IG3) depending on the stated field of study.}
    \label{tab:descriptives.groups}
\end{table}

\subsection{Study Design and Procedure}\label{subsec:design}
The study employed a between-subjects design with a $2 \times 2$ factorial structure. Each participant was randomly assigned to one of four study conditions. All participants were individually presented with the same multimedia learning unit, which consisted of complementary text and image elements that provided non-redundant information. This baseline unit was identical for all groups. The participants' visual behaviour was recorded using a Tobii Pro Nano eye tracker during the learning unit. A nine-point calibration was performed immediately prior to the start of the learning unit to ensure data accuracy.

\begin{figure}
    \centering
    \includegraphics[width=\linewidth]{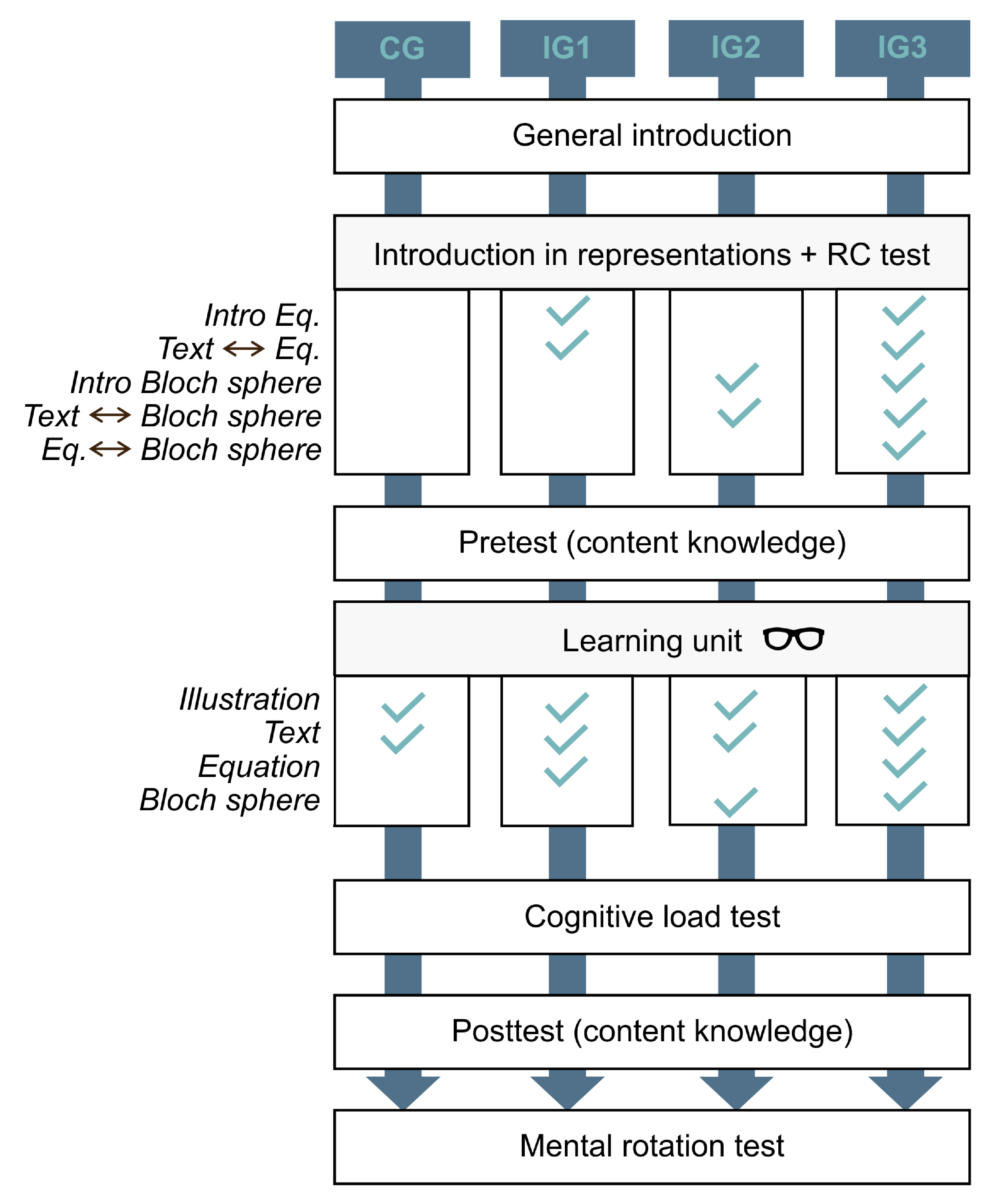}
    \caption{Illustration of the individual stages of the study four the four study conditions CG, IG1, IG2, and IG3. \textit{Note:} Eq., Equation; RC, representational competence.}
    \label{fig:studydesign}
\end{figure}

Two factors were manipulated:

\begin{enumerate}
    \item The presence or absence of an additional graphic-geometric representation (Bloch sphere) that provided redundant information to the text (factor 1: graphic-geometric representation present vs. absent).
    \item The presence or absence of an additional symbolic-mathematical representation (equation) that was also informationally redundant to the text (factor 2: symbolic-mathematical representation present vs. absent).
\end{enumerate}
This design resulted in four experimental groups:

\begin{itemize}
    \item A control group (CG) that received only the baseline multimedia unit without any additional redundant representations,
    \item Intervention Group 1 (IG1), which received the baseline unit plus a graphic-geometric representation,
    \item Intervention Group 2 (IG2), which received the baseline unit plus a symbolic-mathematical representation, and
    \item Intervention Group 3 (IG3), which received the baseline unit plus both the additional graphic-geometrical and the symbolic-mathematical representation.
\end{itemize}
The entire study was conducted through digital means on a computer. The study procedure is described in Figure \ref{fig:studydesign}. In the following paragraphs, we will elucidate the individual stages and materials used in more detail.

\subsection{Materials}
Participants were first given an overview of the basic principles of physics as they relate to light, including a description of the properties of photons. In this regard, the authors designed and recorded a video for use in this study. The participants were permitted to pause, rewind, and fast-forward the video as often as they desired. The video itself did not make any reference to the Dirac formalism or the Bloch sphere. Similar to the first introductory video, the participants were presented with another pre-recorded video outlining the components of the \ac{mzi}. This introduction encompassed the identification of each component and a description of its function within the interferometer. 

\begin{figure}[ht]
    \centering
    \includegraphics[width=\linewidth]{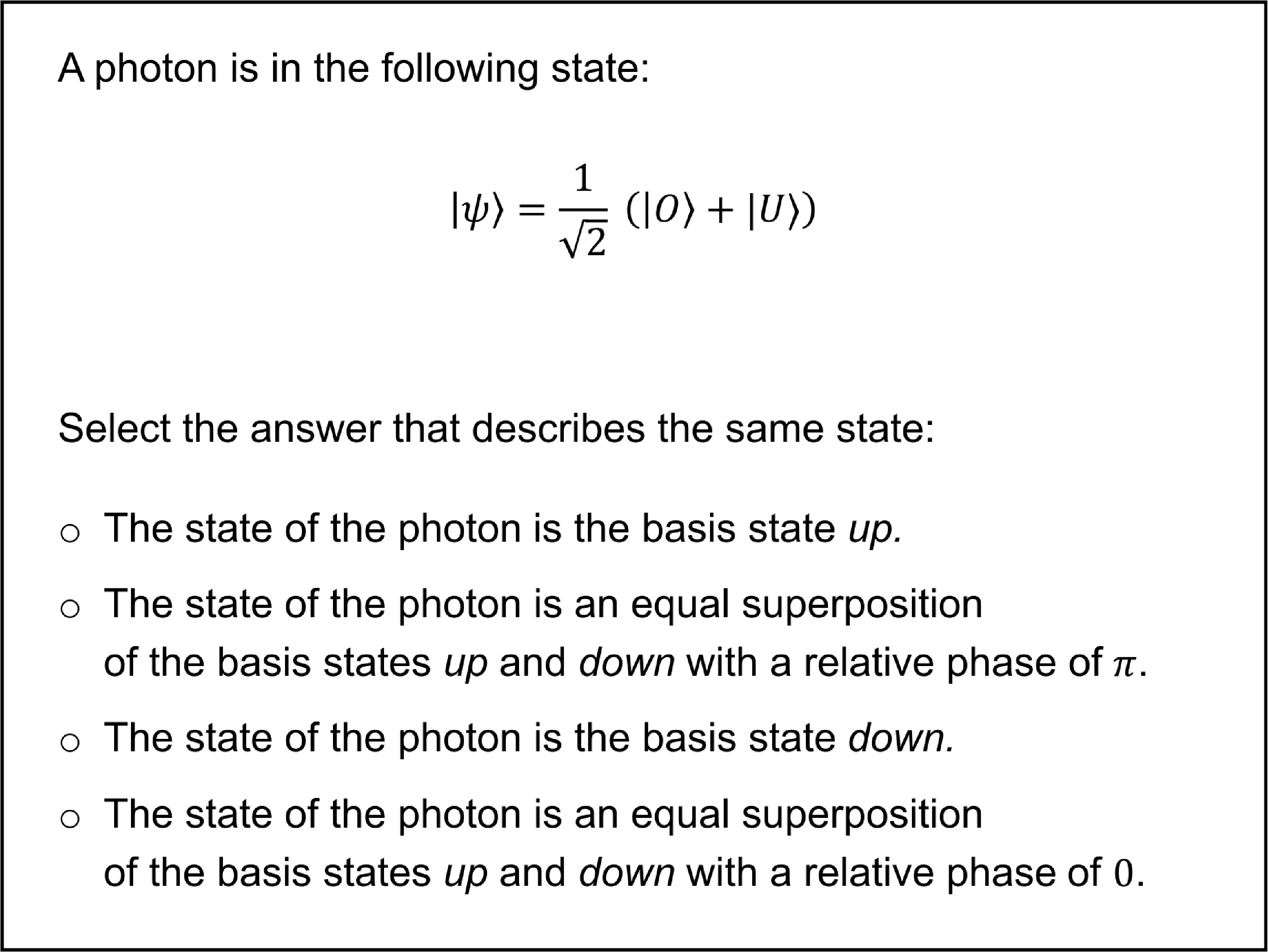}
    \caption{Example item of the representational competence test for the translation between text and equation. Analogous items were used for the translation between text and Bloch sphere and equation and Bloch sphere.}
    \label{fig:RC_example}
\end{figure}

After a general introduction to the subject, each participant was introduced to the external representations specific to their respective group. A brief introductory video was prepared for the Dirac formalism and the Bloch sphere, respectively, in which the method for describing a photon state with the respective external representation was outlined. As in the general introduction, participants were allowed to pause, rewind, and fast-forward as often as they wanted. Participants were instructed to move on to the test phase at their own discretion, ideally after feeling confident in their understanding of the external representation. In the representational competence test, the students were presented with a specific photon state represented in a given external representation and were asked to select the corresponding state in a sample of four presented in another external representation. Depending on the condition assigned, participants worked on different versions of the representational competence test (see Figure \ref{fig:studydesign}). 
The control group was not subjected to this phase of the study.
Participants in IG1 completed the test for translations between equation and text and IG2 for translations between Bloch sphere and text.
Participants in IG3 were asked for both translations between equation and text and Bloch sphere and text, as they were introduced to both additional external representations.
In addition, IG3 was tasked with translating directly between equation and Bloch sphere.
For each set of external representations, participants had to solve four equivalent tasks which differed only in the specific state present.
An example task for translating between equation and text is provided in Figure \ref{fig:RC_example}.

As a third stage of the study, the prior content knowledge of the participants was evaluated. A total of five content-related multiple-choice items were selected from a questionnaire developed by \cite{waitzmann_testing_2024} to assess students' use of quantum reasoning. The questionnaire was designed and validated to assess students' understanding of the core ideas of Probability, Superposition, and Interference (PSI) and has been developed specifically for high school and early undergraduate students (e.g., physics students in their first to third semester). Modifications were made to the items to align them with the formulations used in the study. The items and options were presented in a randomised sequence. In addition to solving the items, the students were asked to indicate their level of confidence in answering each item on a six-point Likert scale, ranging from "very unsure" to "very sure."

\begin{figure*}
    \centering
    \includegraphics[width=\linewidth]{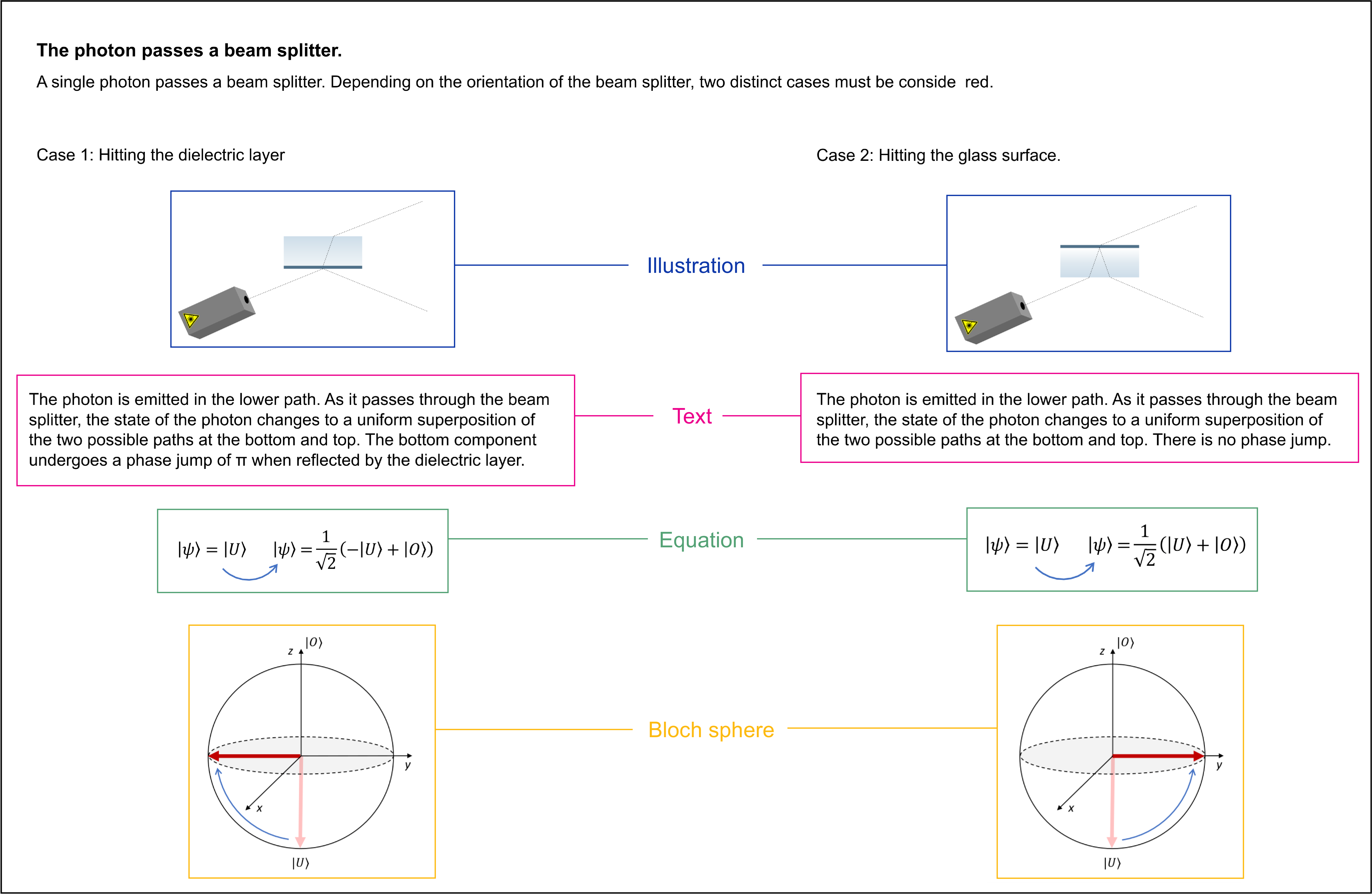}
\caption[Study material for the learning unit]{Study material for the initial stage of the learning unit, as presented to group IG3. In about half of the cases, the placement of the equation and the Bloch sphere was reversed. Depending on the study group, the Dirac formalism and/or the Bloch sphere were omitted. The areas of interest selected for the \ac{et} analysis are highlighted in colour. \par \textit{Note. The text was translated into English for publication, but the study used a German version.}}
    \label{fig:LU1_translated}
\end{figure*}

The learning unit comprised three consecutive stages, corresponding to the scenarios of a photon striking a beam splitter, the addition of a second beam splitter, and the measurement following the second beam splitter. For each stage, participants received a one-page study sheet tailored to their specific study group, with external representations adapted accordingly (see Figure \ref{fig:LU1_translated}). For each stage of the learning unit, participants were asked to answer two to three questions about the content presented in the corresponding material. The students were allowed to switch between the study material and the questions as often as they needed to complete the task. Across the conditions, the participants spent comparable time on the learning unit, with 12.03 minutes ($SD = 5.07$) in CG, 12.08 minutes ($SD = 4.49$) in IG1, 12.89 minutes ($SD = 5.10$) in IG2, and 13.01 minutes ($SD = 5.33$) in IG3.

The learning unit was presented on a 22-inch computer screen with a resolution of $1920\times 1080$ pixels. To capture the visual attention of students during learning, their eye movements were recorded with a stationary Tobii Pro Nano eye tracker. Different types of eye movement (fixations and saccades) were identified using the Identification by Velocity Treshold (I-VT) algorithm with thresholds of $8500^\circ/s^2$ for acceleration and $30^\circ/s$ for velocity. A nine-point calibration was performed before the learning unit for each participant to ensure the accuracy of the detected data. If necessary, the calibration was repeated until it was deemed suitable.

The cognitive load of the participants was evaluated after the completion of the learning unit. For this purpose, the instrument developed by Klepsch et al. \cite{klepsch_development_2017} was chosen, as it provides a validated measure of the different types of cognitive load (\ac{icl}, \ac{ecl}, \ac{gcl}) in a scale from 1 to 7 (disagree or fully agree). The test instrument used is developed and validated in the same language as that of the study participants (German), which was beneficial in its application. It consists of eight items designed to assess cognitive load during the learning unit on the basis of statements. Moreover, compared to other instruments, such as the one proposed by Leppink et al. \cite{Leppink.2013}, the items were found to align best with the learning context of our study. Subsequently, the content knowledge test based on \cite{waitzmann_wirkung_2023} was conducted as a post-test. The test was identical to the one administered as a pretest, with the exception of a randomised order of items and answer options. The capacity for mental rotation was evaluated through the administration of the RCube-Vis test, as proposed by \cite{Fehringer.2020}.

\subsection{Data Analysis}\label{subsec:analysis}

Concerning RQ 1, we analysed the performance and cognitive load of the intervention groups IG1 and IG2 compared to the control group CG.
The performance of each participant was measured in terms of the proportion of correctly solved items, both before and after the learning unit.
The cognitive load was calculated on the basis of subjective ratings according to the dimensions of \ac{ecl}, \ac{icl}, and \ac{gcl}.
To investigate possible differences in performance and cognitive load between the study conditions, we performed a multiple linear regression for each outcome measure, including the pretest accuracy and condition as independent variables.

In order to address RQ2, we also included IG3, receiving both additional external representations, in the respective multiple linear regressions for performance and cognitive load measures.
To establish a linear relationship between each outcome and the condition variable, we transformed the four conditions (CG, IG1, IG2, and IG3) into dummy variables in ascending order according to their average scores on the respective outcome measure.
As representational competence and mental rotation ability were considered potential influencing factors a priori, we subsequently analysed both variables to determine correlations with participants' performance and cognitive load.
To this end, representational competence was defined as the proportion of correct responses on the representational competence test. The mean log-time for correct responses on the mental rotation test was used as a measure of students' mental rotation speed. In line with previous works \cite{jansen2013mental, Fehringer2021} participants with less than 70\% correct answers were excluded from the analysis. In doing so, we ensure a reasonable level of accuracy to derive valid information about spatial ability from the time measure.
Scatterplots were created to illustrate the relationship between the variables and each of the outcome measures.
In order to enhance the robustness of the subsequent statistical analyses, the multiple linear regressions were extended to include the respective variable where feasible.

Third, to answer RQ 3, we performed an analysis of the visual behaviour exhibited by the students within the learning unit. In line with comparable studies in the research field, the areas of interest (AOIs) were designated for each external representation included in the learning unit, depending on the condition \cite{Alemdag.2018}. In the maximum case of condition $IG_3$, each slide of the learning unit comprised four pairs of AOIs, associated with the illustration, the text, the equation, and the Bloch sphere (see \ref{fig:LU1_translated}). We considered transitions between two AOIs of different external representations \cite{Alemdag.2018}, while transitions between the both AOIs for one representation type were omitted. The raw data was detected using the software Tobii Pro Lab, and Python was employed for the identification of transitions, defined as shifts of fixations from one AOI to another. For this purpose, only fixations within the predefined AOIs were taken into account.
The total number of transitions made by the students within the learning unit was analysed using a one-way analysis \ac{anova} with the condition (CG, IG1, IG2, IG3) as independent variable. Moreover, to gain further insight into the distribution of transitions contingent on the specific external representations incorporated into the material in the intervention groups, we conducted an unpaired-sample t-test to compare the relative number of transitions from and to the equation for IG1 and from and to the Bloch sphere for IG2. Similarly, we conducted a paired-samples t-test to compare the relative number of transitions for the two additional external representations in IG3. All statistical analyses were performed with RStudio, version 2023.06.0. Unless otherwise stated, the prerequisites for the respective statistical procedure were verified and found to be satisfied. 

\section{Results}\label{subsec:results}

\subsection{Learning Effectiveness}

\begin{figure}
    \centering
    \includegraphics[width=\linewidth]{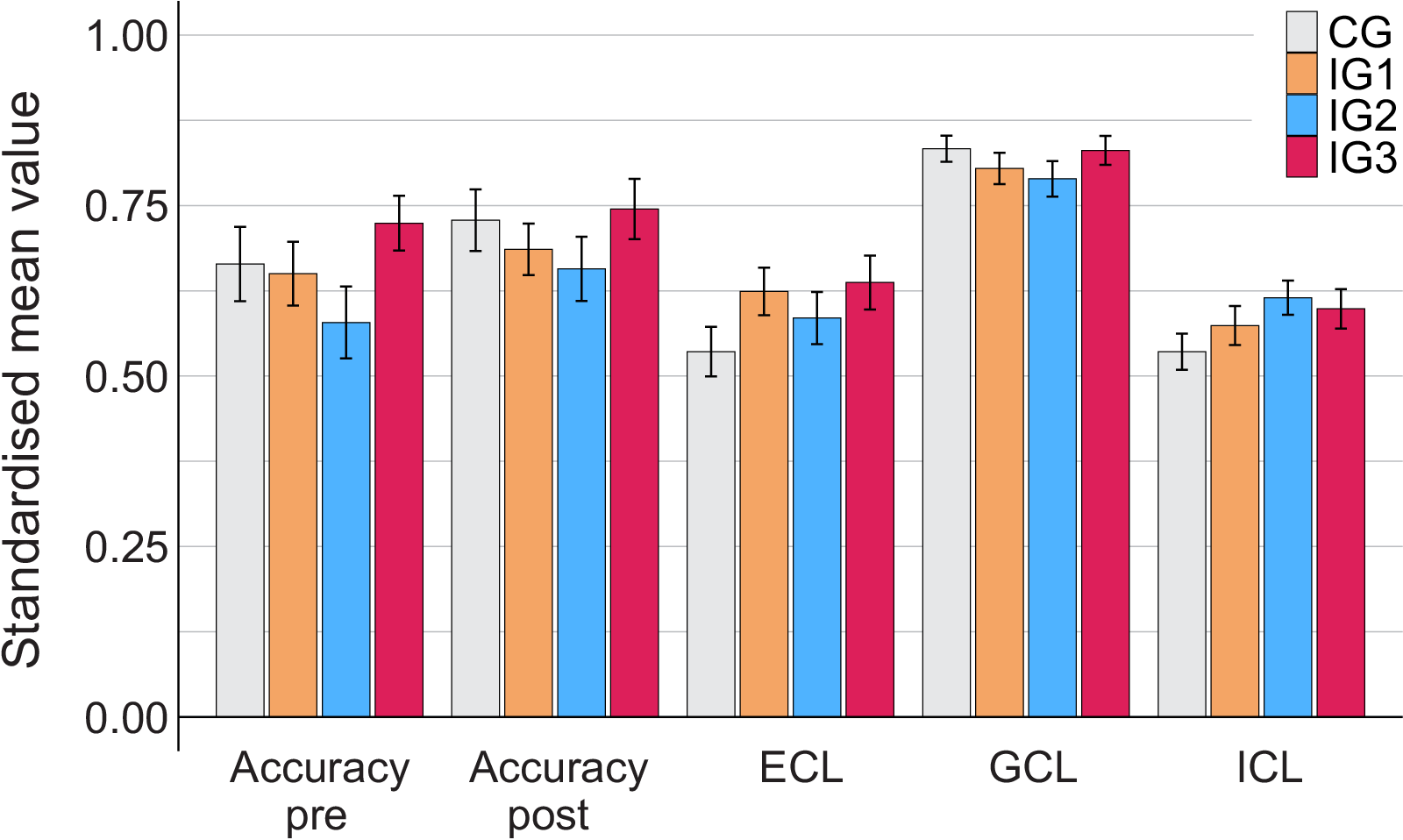}
    \caption{Standardized mean values for the accuracy pre, the accuracy post, \ac{ecl}, \ac{gcl}, and \ac{icl} for the study conditions CG, IG1 and IG2 (28 participants each) and IG3 (29 participants). The error bars represent one standard error.}
    \label{fig:descriptives}
\end{figure}

An overview of the descriptive results for the pretest accuracy, posttest accuracy, and the cognitive load, in terms of \ac{ecl}, \ac{icl}, and \ac{gcl}, is presented in Figure \ref{fig:descriptives} for each of the four conditions involved in the study.
To identify potential differences in student learning across the four conditions, we performed a multiple linear regression analysis for each outcome measure, including the condition and the pretest accuracy as independent variables. The results indicated an overall effect for the accuracy post ($F(4,108)=14.2,~p<.001^{\ast\ast\ast},~R^2=0.345,~R^2_{adj}=0.320$) and the \ac{icl} ($F(4,108)=4.525,~p<.001^{\ast\ast\ast},~R^2=0.144,~R^2_{adj}=0.112$). In contrast, no significant overall effect could be identified for the \ac{ecl} ($F(4,108)=1.165
,~p=.33,~R^2=0.041,~R^2_{adj}=0.006$) and the \ac{gcl} ($F(4,108)=1.547
,~p=.19,~R^2=0.054,R^2_{adj}=0.019$). The results for each independent variable in the statistically significant outcomes of the accuracy post and \ac{icl} visualized in Figure \ref{fig:descriptives} are presented in Table \ref{tab:reg.base}.

\begin{table}[htbp]
    \renewcommand{\arraystretch}{1.2}
    \begin{tabularx}{\linewidth}{l>{\centering\arraybackslash}X>{\centering\arraybackslash}X>{\centering\arraybackslash}X>{\centering\arraybackslash}X}
         \hline
         &$\bm{\beta}$ & \textbf{\textit{SE}}& $\textbf{\textit{t}}$ & \textbf{\textit{p}}\\
         \hline
         \multicolumn{5}{l}{\textbf{Accuracy post}}\\
         Intercept (IG2) & 1.798 & 0.273 & 6.598 & $<.001^{\ast\ast\ast}$ \\
         CG & 0.137 & 0.257 & 0.532 & .60 \\
         IG1 & -0.041 & 0.256 & -0.159 & .87 \\
         IG3 & 0.064 & 0.258 & 0.248 & .80\\
         Accuracy pre & 0.514 & 0.071 & 7.284 & $<.001^{\ast\ast\ast}$ \\
         \multicolumn{5}{l}{\textbf{ICL}}\\
         Intercept (CG) & 4.601 & 0.299 & 15.365 & $<.001^{\ast\ast\ast}$ \\ 
         IG1 & 0.250 & 0.258 & 0.966 & .34 \\
         IG2 & 0.444 & 0.260 & 1.706 & .09 \\
         IG3 & 0.516 & 0.257 & 2.010 & $.05^\ast$ \\
         Accuracy pre & -0.256 & 0.071 & -3,587 & $<.001^{\ast\ast\ast}$ \\ 
         \hline
    \end{tabularx}
    \caption{Individual results for the coefficients of the conditions (CG, IG1, IG2, and IG3) and the pretest accuracy (accuracy pre) of the multiple linear regression for the outcome measures of accuracy post, as well as the \ac{icl}. $^\ast p<.05,~^{\ast\ast\ast} p<.001$}.
    \label{tab:reg.base}
\end{table}

\begin{figure}[ht!]
\captionsetup[subfigure]{justification=raggedright, singlelinecheck=false} % Subfigure Captions linksbündig
    \begin{subfigure}{\linewidth}
        \caption{}
        \centering
        \includegraphics[width=0.65\textwidth]{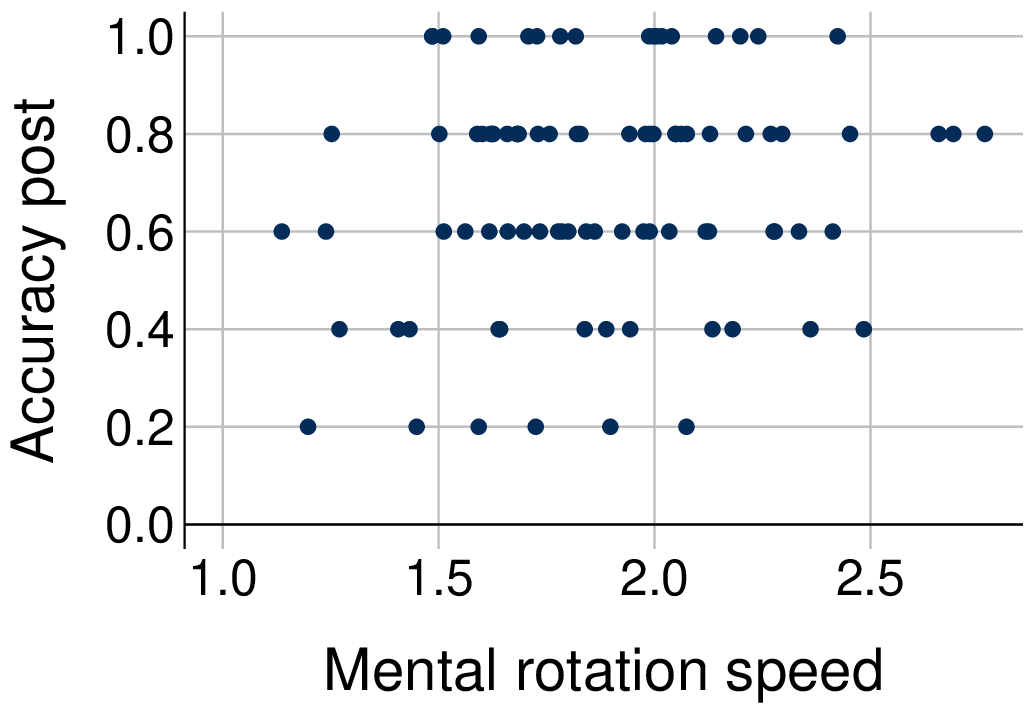}
        \label{fig:MR.acc}
    \end{subfigure}
    \begin{subfigure}{\linewidth}
        \caption{}
        \centering
        \includegraphics[width=0.65\textwidth]{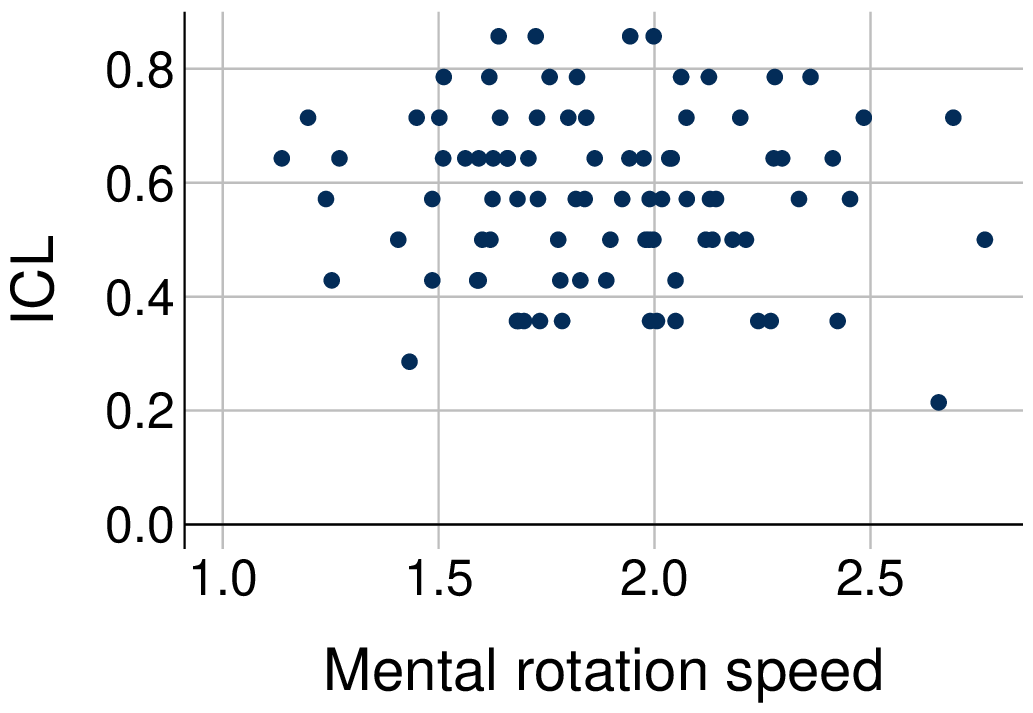}
        \label{fig:MR.ICL}
    \end{subfigure}
    \begin{subfigure}{\linewidth}
        \caption{}
        \centering
        \includegraphics[width=0.65\textwidth]{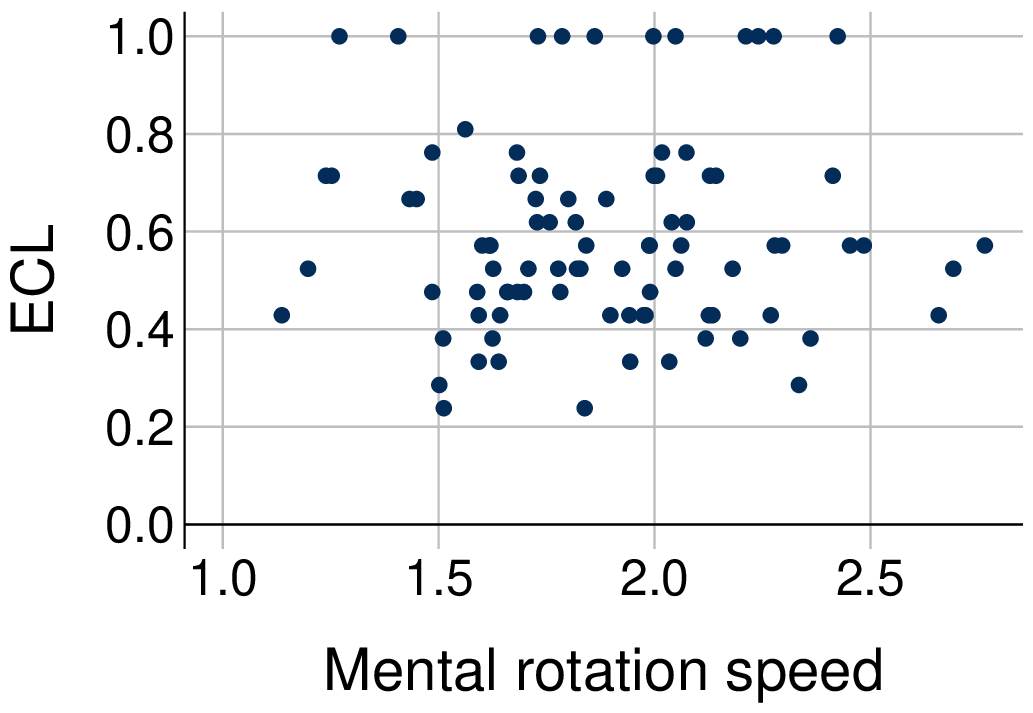}
        \label{fig:MR.ECL}
    \end{subfigure}
    \begin{subfigure}{\linewidth}
        \caption{}
        \centering
        \includegraphics[width=0.65\textwidth]{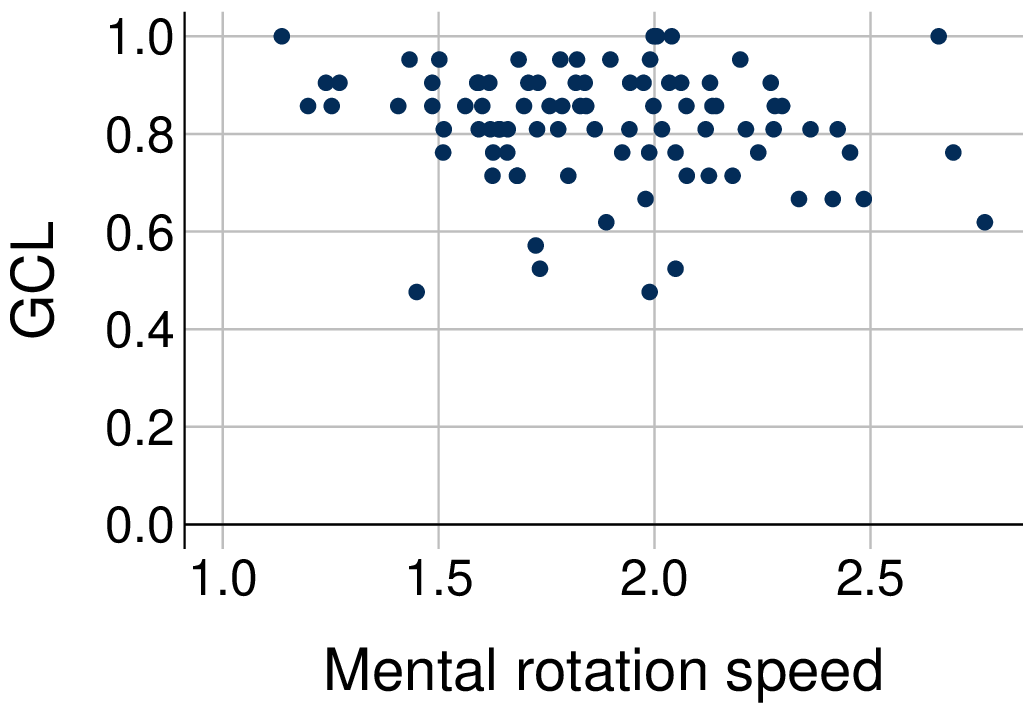}
        \label{fig:MR.GCL}
    \end{subfigure}
     \caption{Scatterplots of the mental rotation ability and (a) the accuracy post (b) the \ac{icl} (c) the \ac{ecl}, and (d) the \ac{gcl} based on the data from 44 participants from IG2 and IG3.}
    \label{fig:Scatter.MR}
\end{figure}

In order to increase the robustness of the previous analyses, we analysed the effect of participants' representational competence in the external representations relevant for the respective intervention group, as well as their mental rotation ability for participants learning with the Bloch sphere.
The findings revealed that the participants demonstrated notably strong performance in the representational competence test. Based on the 49 data sets available for IG2 and IG3 ($M=0.911,~SD=0.167)$, it was observed that $71.43\%$ of the participants attained the maximum score, indicating a high level of proficiency in the external representations provided.
Due to the ceiling effect, the data proved to be unsuitable for identifying potential correlations.
Furthermore, we conducted scatterplots to illustrate the relationship between the mental rotation ability of participants in IG2 and IG3, learning with the Bloch sphere, and each of the outcome measures (see Figure \ref{fig:Scatter.MR}). 

To analyse possible correlations between mental rotation ability and learning outcomes when learning with the Bloch sphere, we performed an extended multiple linear regression for each outcome measure, i.e. accuracy post, \ac{icl}, \ac{ecl} and \ac{gcl}, based on the data of the 44 participants included in the analysis. In doing so, we included the learners' average log time for correct answers in the R-Cube-Vis test as an additional independent variable to the pretest accuracy. 
The analysis yielded a significant overall effect for the accuracy post ($F(2,41)=18.01,~p<.001^{\ast\ast\ast},~R^2=0.468,~R^2_{adj}=0.442$) and \ac{icl} ($F(2,41)=3.356,~p=.04^\ast,~R^2=0.141,~R^2_{adj}=0.099$). However, no significant correlation was identified between mental rotation ability scores and the precision of either of the two outcome measures accuracy post ($\beta=0.081,~SE=0.087,~t=0.924,~p=.36$) and \ac{icl} ($\beta=-0.083,~SE=0.069,~t=-1.207,~p=.23$) (see Table \ref{tab:reg.mr}). Similarly to the basis regression, the overall effect for the outcomes of \ac{ecl} and \ac{gcl} could not be determined to be statistically significant (\ac{ecl}: $F(2,41)=0.411,~p=.67,~R^2=0.020,R^2_{adj}=-0.028$, \ac{gcl}: $F(2,41)=2.516,~p=.09,~R^2=0.109,~R^2_{adj}=0.066$).

\begin{table}[htbp]
    \renewcommand{\arraystretch}{1.2}
    \begin{tabularx}{\linewidth}{l>{\centering\arraybackslash}X>{\centering\arraybackslash}X>{\centering\arraybackslash}X>{\centering\arraybackslash}X}
         \hline
         &$\bm{\beta}$ & \textbf{\textit{SE}}& $\textbf{\textit{t}}$ & \textbf{\textit{p}}\\
         \hline
         \multicolumn{5}{l}{\textbf{Accuracy post}}\\
            Intercept & $0.137$ & $0.176$ & $0.778$ & $.44$\\
            Accuracy pre & $0.617$ & $0.105$ & $5.902$ & $<.001^{\ast\ast\ast}$\\
            MR speed & $0.081$ & $0.087$ & $0.924$ & $.36$\\
         \multicolumn{5}{l}{\textbf{ICL}} \\
            Intercept & $0.874$ & $0.139$ & $6.302$ & $<.001^{\ast\ast\ast}$\\
            Accuracy pre & $-0.186$ & $0.082$ & $-2.259$ & $.029^\ast$\\
            MR speed & $-0.083$ & $0.069$ & $-1.207$ & $.23$\\
         \hline
    \end{tabularx}
    \caption{Summary of regression coefficients of the multiple linear regression for Accuracy pre and MR speed (mental rotation speed) for the outcome measures Accuracy post and ICL, for which significant overall effects were found. No significant effects of mental rotation speed were observed. $^\ast p<.05,~^{\ast\ast\ast} p<.001$}
    \label{tab:reg.mr}
\end{table}

\subsection{Visual Behaviour}

The descriptive results for the total number of transitions $k_{tot}$ are presented in Table \ref{tab:transitions.total}. The one-way \ac{anova} with the condition (CG, IG1, IG2, IG3) as the independent variable and the total number of transitions as the dependent variable yielded a significant overall effect $F(3,94)=8.802,~p<.001^{\ast\ast\ast}$. The results of the subsequent pairwise t-tests with Bonferroni correction are presented in Table \ref{tab:transitions.pairwise}, including the t-value $t$, degree of freedom $df$, the Bonferroni corrected p-value $p_{adj}$ and the effect size Cohen's $d$ for each t-test.

\begin{table}[htbp]
    \renewcommand{\arraystretch}{1.2}
    \begin{tabularx}{\linewidth}{l>{\centering\arraybackslash}X>{\centering\arraybackslash}X>{\centering\arraybackslash}X}
        \hline
        \textbf{Condition} & \textbf{\textit{N}}& $\bm{k_{tot}}$ & \textbf{\textit{SD}}\\
        \hline
        CG   & 24 & 60.58  & 41.38 \\
        IG1  & 23 & 79.22  & 39.52 \\
        IG2  & 22 & 107.59 & 58.07 \\
        IG3  & 25 & 122.72 & 44.48 \\
        \hline
    \end{tabularx}
    \caption{Descriptive data of the number of participants $N$, the mean total number of transitions $k_{tot}$ and the standard deviation $SD$ for each of the four conditions.}
    \label{tab:transitions.total}
\end{table}

\begin{table}[htbp]
    \renewcommand{\arraystretch}{1.2}
    \begin{tabularx}{\linewidth}{l>{\centering\arraybackslash}X>{\centering\arraybackslash}X>{\centering\arraybackslash}X>{\centering\arraybackslash}X}
        \hline
        \textbf{Conditions} & \textbf{\textit{t}} & \textbf{\textit{df}} & $\bm{p_{adj}}$ & \textbf{\textit{d}}\\
        \hline
        IG1 vs. CG  & $1.58$ & 45 & $.73$ & $0.46$\\
        IG2 vs. CG  & $3.18$ & 44 & $.02^{\ast}$ & $0.94$\\
        IG3 vs. CG  & $5.06$ & 47 & $<.001^{\ast\ast\ast}$ & $1.45$\\
        \hline
        IG2 vs. IG1 & $1.92$ & 43 & $.37$ & $0.57$\\
        IG3 vs. IG1 & $3.57$ & 46 & $.01^{\ast\ast}$ & $1.03$\\
        \hline
        IG3 vs. IG2 & $1.01$ & 45 & $1.00$ & $0.30$\\
        \hline
    \end{tabularx}
    \caption{Results of the pairwise t-tests for the total number of transitions $k_{tot}$ between the four conditions CG, IG1, IG2, and IG3 with Bonferroni corrected p-values $p_{adj}$ and effect size Cohen's \textit{d}. $^{\ast\ast} p<.01$, $^{\ast\ast\ast} p<.001$.}
\label{tab:transitions.pairwise}
\end{table}

We calculated the relative number of transitions from and to the equation $k_{rel, eq}$ for IG1 and the Bloch sphere $k_{rel,b}$ for IG2. Moreover, we calculated the relative number of transitions for the two additional external representations in IG3. The results are presented in Table \ref{tab:transitions.rel}. Since IG3 was presented with the equation and the Bloch sphere at the same time, transitions between the additional external representations are included in both $k_{rel,eq}$ and $k_{rel,b}$. Therefore, for IG3, the sum of $k_{rel,eq}$ and $k_{rel,b}$ is not necessarily equal to the total number of transitions made by the participants. The unpaired t-test for the intervention groups IG1 and IG2 did not yield statistically significant differences for $k_{rel, eq}$ and $k_{rel, b}$ ($t(44)=1.74,~p=.09$). Furthermore, the corresponding paired sample t-test for the intervention group IG3 did not reveal significant differences ($t(24)=-0.81,~p=.42$).

\begin{table}[htbp]
    \renewcommand{\arraystretch}{1.2}
    \begin{tabularx}{\linewidth}{l>{\centering\arraybackslash}X>{\centering\arraybackslash}X>{\centering\arraybackslash}X>{\centering\arraybackslash}X}
        \hline
        \textbf{Condition} & \textbf{\textit{N}}& $\bm{k_{rel,eq}}$ & $\bm{k_{rel,b}}$ & \textbf{\textit{SD}}\\
        \hline
        IG1  & 23 & 0.54  & $-$ &0.17\\
        IG2  & 22 & $-$ & 0.46 &0.15\\
        IG3  & 25 & 0.48 & 0.52 & 0.14\\
        \hline
    \end{tabularx}
    \caption{Overview over the number of participants \textit{N}, the relative number of transitions from and to the equation $k_{rel,eq}$ and the Bloch sphere $k_{rel,b}$, and the standard deviation \textit{SD} for each of the intervention groups.}
    \label{tab:transitions.rel}
\end{table}

\section{Discussion}

The objective of this study was to investigate the effects of extending a multimedia learning unit with additional symbolic external representation, specifically equations expressed in the Dirac formalism, or a graphical representation, namely the Bloch sphere, on students learning of quantum properties. In particular, both additional external representations are redundant in terms of the relevant information content, given the multimedia basis of the text and illustration. 

\subsection{Learning Effectiveness}
In regard to RQ1, no significant effects on students' content knowledge could be detected when learning with the additional symbolic external representation or when provided with the additional graphical external representation, in comparison to the basis multimedia unit.  Contrary to previous results and assumptions \cite{rexigel_more_2024}, the provision of more informational redundant external representations was not associated with better learning outcomes. Similarly, students enrolled in IG1, who received additional instruction through equations, and students enrolled in IG2, who received additional instruction through the Bloch sphere, exhibited comparable cognitive load (as indicated by \ac{icl}, \ac{ecl} and \ac{gcl}) to that observed in the CG, who were provided with the fundamental multimedia setting alone. Consequently, with regard to RQ1, providing students with an additional symbolic or graphical external representation did not result in discernible improvements in content knowledge or cognitive load. Therefore, the findings of this study do not support the conclusions of previous research in other contexts \cite[e.g.,][]{ott_multiple_2018}, proposing a possible advantage of learning with \ac{mer} with shared essential information. According to the \ac{deft} framework, \ac{mer} with shared essential information have the potential to improve learning outcomes by prompting different cognitive processes or providing the opportunity to choose the external representation most appropriate for learning \cite{ainsworth_deft_2006}, especially in settings of more than two external representations \cite{rexigel_more_2024}. 
Despite the fact that the vast majority of the participants demonstrated a high level of proficiency in using the external representations provided, as evidenced by their notable achievements in the representational competence test, the findings suggest that the learners in this study did not realise the potential benefits of learning with multiple informational redundant external representations. 
However, the incorporation of additional redundant representations did neither result in a decline in students' learning outcomes, as measured by content knowledge and cognitive load. Consequently, the analysed outcomes do not provide clear support for the preceding research that indicated positive effects of the incorporation of redundant external representations in students' learning \cite[e.g,][]{ainsworth_deft_2006, rexigel_more_2024}, or negative effects \cite{Kalyuga.2021}. One explanation for the absence of observed effects might be found in the measurement tools used to assess content knowledge and cognitive load. While both instruments employed in this study are validated, it is possible that they lack sufficient sensitivity to detect differences in the given context. An alternative explanation may be found in a more intricate interaction of student characteristics and the effect of redundant external representations on learning quantum properties.

In order to account for possible influencing factors of the findings, especially for students learning with the additional Bloch sphere, data were collected about students' representational competence and mental rotation ability.
The participants obtained commendable results in the representational competence test, suggesting a high level of proficiency in the use of the respective external representations that were presented. Consequently, it is not reasonable to assume that any potential benefits of the additional redundant representation would be outweighed by inadequate representational competencies.
Considering the potential impact of mental rotation ability on the efficacy of learning using the Bloch sphere, the absence of a significant correlation between mental rotation ability and either performance or cognitive load indicates that, if such effects exist, they are overshadowed by the influence of prior knowledge on the learning outcome.
A possible explanation for the lack of effects of mental rotation ability on learning with the Bloch sphere might also be that the learning tasks used in our study did not require continuous or complex spatial transformations. It is possible that the learners have relied more on conceptual or symbolic strategies. In addition, didactic support is provided by the clear display of the directions of vector rotation. This may have reduced the need for high mental rotation ability, thus diminishing the predictive power of individual differences in spatial ability in this context.
Nevertheless, given the limited number of participants, particularly with regard to their mental rotation ability ($k = 45$), and the consequent limited statistical power, it is possible that some statistically significant results may have been missed.

With regard to RQ2, we also investigated the potential impact of incorporating both informationally redundant external representations into the multimedia learning unit (IG3). As in the intervention groups IG1 and IG2, who received one of the two additional external representations, the presentation of the equation and the Bloch sphere did not result in an improved knowledge of the content. However, students who learnt with the maximum combination of four external representations demonstrated an increased \ac{icl}. Following the \ac{clt} \cite{sweller_cognitive_2019} and the \ac{ctml} \cite{Mayer.2021e} the results imply that the addition of \ac{mer} with informational redundancy leads to enhanced element interactivity and, correspondingly, enhanced essential processing. According to Mayer's definition, learning with both additional external representations is associated with greater cognitive processing in order to represent the essential information in working memory \cite{Mayer.2021e}. As IG3 did not result in an enhancement of content knowledge, the findings indicate that the provision of supplementary external representations induced students to perceive the learning content as more complex and challenging, with no evident advantages in content knowledge. 

In consideration of RQ1 and RQ2, the provision of an additional informationally redundant symbolic or/and graphical external representation was not associated with advanced learning outcomes.

\subsection{Visual Behaviour and Learning Effectiveness}

The analysis of the learning outcome in relation to the presence of additional informationally redundant external representations indicated that there was no discernible impact on students' content knowledge when learning with a multimedia learning unit. However, the analysis of the cognitive load of the students when learning indicated that the participants in IG3, who received the maximum set of four external representations, experienced a higher level of \ac{icl} than the participants in CG, who learnt in the basic multimedia setting with two complementary external representations. This suggests that, although there were no differences in final content knowledge, the additional external representations may have prompted the use of different learning strategies. To gain insight into the learning processes employed according to the study condition, we conducted an analysis of the visual behaviour exhibited by students during the learning process. In line with previous research, we analysed the total and relative number of transitions between external representations as an indicator of attempted integration processes \cite{Coskun.2022}.

A higher number of transitions between external representations can be related to students' learning outcome in different ways.
Research has indicated that an increased number of transitions is associated with better understanding and transfer performance when learning with \ac{mer} \cite[e.g.,][]{kragten2015students, o2014learning}. In other contexts, frequent transitions between external representations can also be indicative of processing difficulties and have a detrimental effect on learning success \cite{Coskun.2022, Alemdag.2018}. Consequently, a high number of transitions may reflect successful integration processes or processing difficulties \cite{Coskun.2022}. It is therefore essential to consider both the instructional design of external representations, individual learner characteristics and the learning outcome when interpreting transition frequency as a proxy for learning effectiveness.

The statistical analysis indicates that students demonstrated a higher total number of transitions  between the external representations presented when the Bloch sphere was provided as an additional graphical external representation in the learning material. This was observed not only in IG2, who learnt only with the additional Bloch sphere, but also in IG3, who learnt both with the additional equation in the Dirac notation and the Bloch sphere. Given that an additional external representation, even if it does not provide any new information content, represents a further processing source, it is reasonable to expect an increase in integrations with more representations. However, the results indicate that the enhancement in transitions is only related to the presentation of the additional graphical external representation, not the symbolic one.
Although the basic multimedia unit comprised a symbolic external representation (text) and a graphical one (illustration), the essential information about the quantum state in different phases when passing the \ac{mzi} is conveyed by the text. 
Moreover, the text constitutes the informationally redundant reference representation.
Therefore, redundant information is still presented in the homogeneous combination of text and equation for IG1.
In contrast, the incorporation of the Bloch sphere results in the presentation of redundant information in the heterogeneous combination of text and Bloch sphere for IG2.
Consequently, the increase in attempted integration behaviour exhibited by participants learning with the Bloch sphere is consistent with the findings of previous research \cite[e.g.,][]{ott_multiple_2018}.
Here, a higher number of transitions was observed in heterogeneous combinations of \ac{mer} compared to homogeneous combinations comprising only symbolic external representations \cite{ott_multiple_2018}. 

In line with the previous considerations, an increase in the number of transitions was not only observed when comparing CG, who received the basic multimedia setting, with IG2 or IG3, who received either the additional Bloch sphere (IG2) or additional equations using the Dirac formalism and the Bloch sphere (IG3). An increase in transitions was also detected when IG1, which received additional equations, was compared to IG3, where the Bloch sphere was added to the IG1 setting. 
While in IG1 the essential information regarding the basis state itself is provided by a homogeneous combination of text and equation, redundantly, the additional Bloch sphere in IG3 results in a presentation of redundant information across the heterogeneous combination of text, equation and Bloch sphere.
Once more, the presentation of a heterogeneous combination of redundant representations, in this case given by text, equation and Bloch sphere, is associated with an increase in attempted integration processes, in line with previous research \cite{ott_multiple_2018}.
It can thus be concluded that in the present study the Bloch sphere plays a central role in the learning process, encouraging learners to proactively seek to connect information from different sources by facilitating the presentation of redundant information in heterogeneous external representations. 

Interestingly, these increased transitions were not limited to the Bloch sphere itself with the other external representations presented, as indicated by the subsequent analysis of transitions to and from the additional external representation. When comparing IG1, receiving additional equations and IG2, receiving the additional Bloch sphere, similar relative numbers of transitions were found for each of the additional external representations. Similar findings were observed when the relative number of transitions from and to the equation and the Bloch sphere in group IG3 was considered. As a result, the provision of the Bloch sphere appears to encourage an increased level of attempted integration that encompasses all of the learning material.
This could indicate an attempt to establish connections between the various external representations with the aim of developing a more comprehensive understanding. Although the integration of diverse external representations can be advantageous \cite{rau_conditions_2017}, the additional cognitive effort required did not result in improved learning outcomes. Consequently, the approach was not efficient in the context of this study. 
We found ceiling effects in the representational competence test, conducted previous to the learning unit. This suggests that learners were well-versed in handling the external representations used in the study. Consequently, the observed increase in transitions is unlikely to result from insufficient representational competence.

Another possible explanation for the observed cognitive processing differences might lie in the design of the graphical external representation itself. The Bloch sphere is not only based on icons, the fundamental unit of any graphical external representation \cite{Schnotz.2021}. It also incorporates symbolic elements to signify the fundamental states and the labelling of the axes. Thus, it combines properties of both graphical and symbolic representations, which are partly also found in the other external representations provided. In particular, the Bloch sphere encompasses the presentation of the two basis states in Dirac notation, as also included in the equation. Consequently, the additional equation may be regarded as a logical reference point, as it unifies the symbolic representation of the basis states in terms of the Dirac notation. It can thus be concluded that the promotion of unused cognitive processing may be attributed to the particular characteristics of the Bloch sphere, rather than being a phenomenon inherent to graphical external representations.

\subsection{Limitations and Future Research}

There are some limitations in our study that may serve as a starting point for further research. 
In the current study, the incorporation of a redundant graphical external representation, the Bloch sphere, was found to be associated with less efficient learning processes. 
Despite the lack of detected benefits in terms of content knowledge and cognitive load, it is possible that the test methods employed have failed to identify potential benefits of the Bloch sphere.
For instance, it is conceivable that more profound integration processes may have led to the formation of more robust and connectible schemata, which were not detected by the outcome assessments used.
Nevertheless, the eye-tracking analysis conducted proved to be highly sensitive, uncovering differences that a simple multiple-choice post-test would not have been able to detect. 
It may be advantageous for further research to focus on outcome measurements that are more sensitive, and to extend the scope of immediate performance assessments.
For example, conceptual knowledge could be measured through open-ended explanations or concept-mapping tasks to assess a deeper understanding of the underlying principles.
Transfer effects might be evaluated by examining how well learners apply acquired knowledge to new problems or different contexts. 
Additionally, follow-up tests, such as delayed assessments, could provide insight into the long-term retention and solidity of learning effects.

The \ac{et} analysis conducted may provide a foundation for subsequent fine-grained analyses of students' cognitive processes when learning with \ac{mer} in quantum education.
The present study provides initial insights into the different visual processing of the Dirac formalism and the Bloch sphere in the given context.
Further studies could focus on which elements of the external representations are relevant for the respective visual processing.
For instance, subsequent studies could investigate which parts of the text precede or follow the transition to or from the equation and Bloch sphere.
This approach may facilitate a more precise understanding of the relevant elements of the representations involved in the learning process.

To gain further insight into the generalisability of the findings, more research is required on different combinations of informationally redundant external representations.
In particular, future studies could explore additional graphical external representations commonly used in quantum physics, such as Feynman diagrams \cite{feynman2006qed} or recent external representations such as the Circle notation \cite{johnston_programming_2019, bley_visualizing_2024}.
Investigating these alternatives could help determine whether the observed facilitation of integration behaviour is specific to the Bloch sphere or reflects a more general phenomenon of heterogeneous \ac{mer} with shared information.
At this point, it is unclear whether the different learning strategies associated with the additional symbolic and graphical external representation are a generalisable phenomenon across different types of external representations or whether they are triggered by individual characteristics of the Dirac formalism and the Bloch sphere.
Future research should include different symbolic and graphical external representations to investigate whether the findings can be replicated.

Another limitation of our study is that most of the participants had a \ac{stem} background and were already accustomed to mathematical formulas as external representations in their studies, which may have influenced their perception and processing of these external representations. We did not detect an increased cognitive load associated with their use, which might be explained by the fact that \ac{stem} students are already familiar with this type of external representation from their studies. This familiarity could have mitigated the cognitive demands typically associated with the processing of complex symbolic external representations.

Furthermore, investigating the effects of \ac{mrc} when learning with informationally redundant external representations could be a valuable addition to future research. It could provide deeper insight into how learners choose and use external representations effectively. As diSessa (2004) states, 
\begin{quote}
    "\textit{\ac{mrc} includes the ability to select, produce, and use external representations productively, as well as the ability to critique, modify, and even design entirely new representations."~\cite{disessa_meta-representation_2000}}
\end{quote}

Addressing \ac{mrc} in future studies would allow a more nuanced understanding of the strategies associated with learning with redundant external representations and related learning outcomes.

\section{Conclusion}

This study provides initial insight into the role of redundant external representations in learning fundamental quantum concepts in the context of the \ac{mzi}. It is among the first investigations into the use of \ac{mer} in this domain, particularly with regard to their effects on learning and cognitive processing. Consequently, the findings cannot yet be directly translated into concrete recommendations for teaching. However, one key observation is that adding one or more informationally redundant external representations to multimedia learning materials in the field of quantum properties does not necessarily lead to significant learning gains or losses. 

Nevertheless, the inclusion of graphical-geometric external representations, such as the Bloch sphere, appears to encourage learners to attempt integration between different external representations. This is reflected in an increase in transition behaviour, which, in turn, results in higher intrinsic cognitive load (ICL). These findings align with prior research on \ac{mer}, which suggests that graphical external representations may facilitate cognitive integration, even if this does not directly translate into measurable learning benefits \cite{ott_multiple_2018, Schnotz.2021}. 

Although this study does not yet allow definitive conclusions regarding practical applications, it demonstrates that the choice of external representations significantly influences how learners interact with the material. Further targeted research in quantum physics education with \ac{mer} is therefore warranted.

\subsection{Practical Implication}

The findings provide preliminary insights into how redundant external representations influence learning processes in complex domains such as quantum physics. Although no differences in learning outcomes were detected depending on the number and type of informationally redundant \ac{mer} included, differences in cognitive processing suggest that the design of instructional materials should carefully consider the role of additional external representations. In particular, the inefficient visual behaviour observed when learning with the Bloch sphere suggests that additional scaffolding or targeted cues may be necessary to help learners effectively integrate such external representations.

Key aspects to consider for the design of instructional material, especially in the context of quantum physics:

\begin{enumerate}
    \item \textbf{Strategic integration of redundant external representations}: The use of additional external representations should be approached deliberately, balancing their potential to promote visual integration with their impact on cognitive load \cite{ainsworth_deft_2006, schnotz2014strategy}. 

    \item \textbf{Developing representational competence}: Learning materials should not only support the understanding of individual external representations, but also help learners develop the ability to transition between different formats. Graphical external representations, such as the Bloch sphere, may foster these transitions. While this might not directly enhance content learning, it could contribute to representational fluency by facilitating students ability to connect \ac{mer} efficiently \cite{rau_conditions_2017}.

    \item \textbf{Supporting learners in handling complex external representations}: The benefits of complex graphical external representations, such as the Bloch sphere, may only be fully realised if the learners receive adequate support. Scaffolding approaches, including guided instructions or structured tasks, could be beneficial in helping students navigate and integrate these external representations effectively \cite{Mayer.2021g}.
\end{enumerate}

\begin{acknowledgments}
We would like to thank Prof. Dr. Jürgen Eschner and Prof. Dr. Christoph Becher from the Saarland University for their support in the data collection.
E.R. and J.B. acknowledge the support by the German Federal Ministry of Education and Research (BMBF) through the QuanTUK project.
Furthermore, E.R. acknowledges the support by the European Union via the Interreg Oberrhein programme in the project Quantum Valley Oberrhein (UpQuantVal).
This work was supported by LMUexcellent, funded by the Federal Ministry of Education and Research (BMBF) and the Free State of Bavaria under the Excellence Strategy of the Federal Government and the Länder.
\end{acknowledgments}

\bibliographystyle{apsrev4-2}

\clearpage

\section*{List of acronyms}
\begin{acronym}[ANOVA] % add longest acronym here for alignment
\acro{stem}[STEM]{Science, Technology, Engineering, \& Mathematics}
\acro{deft}[DeFT]{Design, Functions, and Tasks}
\acro{mer}[MERs]{Multiple External Representations}
\acro{qp}[QP]{Quantum Physics}
\acro{cf}[CF]{European Competence Framework}
\acro{itpc}[ITPC]{Integrated Theory of Text and Picture Comprehension}
\acro{ctml}[CTML]{Cognitive Theory of Multimedia Learning}
\acro{clt}[CLT]{Cognitive Load Theory}
\acro{anova}[ANOVA]{Analysis of Variance}
\acro{ecl}[ECL]{Extraneous Cognitive Load}
\acro{icl}[ICL]{Intrinsic Cognitive Load}
\acro{gcl}[GCL]{Germane Cognitive Load}
\acro{et}[ET]{Eye Tracking}
\acro{mrc}[MRC]{Meta-Representational Competencies}
\acro{mzi}[MZI]{Mach-Zehnder Interferometer}
\end{acronym}

\end{document}